\documentstyle[amssymb,aps,prl]{revtex}

\begin{document}
\title{Dissipation, noise and vacuum decay in quantum field theory}
\author{Esteban Calzetta}
\address{Departamento de F\'{\i}sica,\\
Universidad de Buenos Aires, Ciudad Universitaria,\\
1428 Buenos Aires, Argentina}
\author{Albert Roura and Enric Verdaguer \thanks{
Also at Institut de F\'\i sica d'Altes Energies (IFAE), Barcelona, Spain.}}
\address{Departament de F\'{\i}sica Fonamental,\\
Universitat de Barcelona, Av.~Diagonal 647,\\
08028 Barcelona, Spain}
\maketitle

\begin{abstract}
We study the process of vacuum decay in quantum field theory focusing on
the
stochastic aspects of the interaction between long and short-wavelength
modes. This interaction results in a diffusive behavior of the reduced
Wigner function describing the state of the long-wavelength modes, and
thereby to a finite activation rate even at zero temperature. This effect
can make a substantial contribution to the total decay rate.
\end{abstract}


In this Letter we shall investigate how stochastic interactions between
long
and short-wavelength modes affect vacuum decay in scalar quantum field
theory. Two of the present authors have already considered the relevance of
stochasticity in the context of the {\it creation from nothing} of the
Universe \cite{CV}. Our analysis of that problem led to the conclusion that
the noise-induced transition amplitude was actually larger than the usual
quantum estimates \cite{Vilenkin}. However, it remained unclear whether the
relevance of stochasticity for the full decay amplitude was a peculiarity
of
gravitationally bound systems, or rather a generic feature of vacuum decay
in field theory. The results we shall discuss here point quite conclusively
in the second direction. In pursuit of clarity, we shall omit most of the
technical details, which shall be reported in separate publications \cite
{paper1,paper2}

As a simple nongravitational example, let us consider a self-interacting
scalar field $\Phi $ in Minkowski spacetime. The classical action is 
\begin{equation}
S_{ren}\left[ \Phi \right] =-{\frac{1}{2}}\int d^{4}x\left( \partial \Phi
\cdot \partial \Phi +M^{2}\Phi ^{2}-\frac{1}{3}g\Phi ^{3}\right) \text{.}
\label{action}
\end{equation}
Although we keep $\hbar $ explicit we set $c=1$, $M$ has units of $%
lenght^{-1}$, $\Phi $ has units of $M\sqrt{\hbar }$ and $g$ of $M/\sqrt{%
\hbar }$. For simplicity, we shall assume that renormalization has already
been carried out and that Eq. (\ref{action}) is a good description of the
relevant dynamics. This means that the parameters $M^{2}$ and $g$ may well
be renormalization point dependent; in any case, any such dependence will
be
taken as given. This model may be considered as the limiting case of the
class of models studied by Baacke and Kiselev \cite{bk1}, when the coupling
in the quartic self-interaction is very small.

We are concerned with situations where the potential displays a local
minimum, separated from the absolute minimum by a potential barrier. A
system of few degrees of freedom, prepared in a false vacuum state within a
potential well, may decay in essentially two different ways, namely by
tunnel effect, that is, going through the barrier in a classically
forbidden
trajectory, or else, by activation, that is, jumping over the barrier \cite
{Kramers,Langer1}. In systems with few degrees of freedom, there must be an
external agent, typically a thermal source, for activation to be possible.
Activation results from the system being driven by noise originating in the
source.

In either case, the decay probability follows the Arrhenius law $P\sim
Ae^{-B}.$ In the tunnel effect, $B=S_{E}/\hbar $, where $\hbar $ is
Planck's
constant and $S_{E}$ is the action for the trajectory which goes under the
barrier in Euclidean time \cite{Coleman}. In activation, $B=V_{s}/k_{B}T$,
where $k_{B}$ is Boltzmann's constant, $T$ is the temperature, and $V_{s}$
is the height of the free energy barrier measured from the false vacuum
\cite
{HTB}. We can see that activation disappears as $T\rightarrow 0$.

Our thesis is that in field theories there is a phenomenon similar to
activation, even in the absence of an external environment and, most
remarkably, that this simulated activation contributes to vacuum decay
probability even at zero temperature. This phenomenon exists because, while
vacuum decay concerns mainly the long-wavelength modes in the field, these
modes evolve in the environment provided by the short-wavelength ones. Due
to the time dependence of the long-wavelength modes, even if the
short-wavelength modes were initially prepared in their vacuum states,
these
will evolve into coherent superpositions of many particle states. The
energy
to create these particles is provided by the long-wavelength modes. On the
other hand, it is not possible to predict the exact number of particles to
be created. For Bose-Einstein statistics, for example, if $N$ particles are
created in the mean, then the dispersion in this number is of order
$\sqrt{%
N\left( N+1\right) }$, and it is never negligible.

Therefore, we find a dissipative term in the dynamics of the
long-wavelength
modes, representing the energy transfer towards the short-wavelength modes,
but also a stochastic element, related to the fluctuations in the energy
flux. These two terms are related to each other through the
fluctuation-dissipation theorems. We must stress that the presence of one
of
them implies the presence of the other as well. Unlike in Kramers'
activation, this environment is intrinsic to the system. We should point
out
that because of this same reason, we are not allowed to prescribe the
characteristics of noise and dissipation independently of the system
dynamics. This means that it is not possible in general to assume ohmic
dissipation or white noise \cite{Horsthemke}.

More concretely,
tunneling occurs in models where the system may be trapped into a
metastable
state, which is separated from the basin of attraction of the true vacuum
by
a potential barrier. There is a saddle point on this barrier, representing
the critical bubble, and most of the tunneling dynamics is concerned with
motion along the most likely escape path \cite{bbw}, which goes through the
saddle in the direction of steepest descent. It is possible to identify a
few degrees of freedom which parametrize the different configurations on
this path; the remaining (infinite) degrees of freedom describe deviations
from the most likely escape route. In the conventional approaches to
tunneling, the role of these fluctuations is downplayed: they renormalize
the action for the few distinguished parameters, and provide a prefactor
which, after the contributions to the effective potential have been
included
in the exponential, is of order one \cite{prefactor}. We intend to focus on
the back-reaction of the fluctuations away from the most likely escape path
on the quantum dynamics of the critical bubble; for this purpose, we shall
borrow tools from quantum open systems theory, by considering the few
distinguished degrees of freedom as a {\it system} interacting with the
{\it %
environment} provided by the transversal fluctuations.

The technical complexity of the problem increases sharply with the amount
of
information one wishes to retain within the {\it system}. For example, one
may parametrize the most likely escape path following Ref. \cite{vv1}. As
shown in this reference (and earlier in Ref. \cite{Rubakov}), a time
dependent
bubble excites the degrees of freedom in the environment, thus setting up
the kind of dynamical interplay we wish to analyze
(see also Ref. \cite{boy99}).
To be able to focus on the new (dynamical) aspects of the problem, over and
above its geometrical aspects, we shall adopt an intentionally simplified
parametrization of the most likely escape path which retains the essentials
of the physics involved.

Let us return to the scalar field theory above. The potential $V\left[ \phi
\right] =\frac{1}{2}M^{2}\phi ^{2}-\frac{1}{6}g\phi ^{3}$ has a stable
fixed
point at $\phi =0$ and an unstable fixed point at $\phi =\phi
_{s}=2g^{-1}M^{2}.$ The former corresponds to zero energy, and the latter
to $%
E=E_{s}=VM^{2}\phi _{s}^{2}/6$ in a volume $V$. For intermediate energies,
we may have bound and unbound states. They are separated by a potential
barrier, which at zero energy extends from $\phi =0$ to $\phi =\phi
_{exit}=3\phi _{s}/2$. To identify the relevant modes, we observe that, if
we consider fluctuations around the unstable fixed point $\phi _{s}$, then
modes with wavenumber $k>M$ are stable. The relevant modes, which partake
in
the tunneling process, have $k<M$ \cite{dan}. We therefore write the field
as $\Phi =\phi +\varphi ,$ where the first term contains only modes with $%
k<M $, and the second term contains the short wavelengths; $\phi $ shall be
our system.

In other words, the field $\phi $ represents the average of the full field
$%
\Phi $ over volumes of order $M^{-3}$. By construction, $\phi $ is slowly
varying in space; it is technically simplest to handle it as if it were
actually
spatially homogeneous. We shall therefore regard the configurations along
the most likely escape path as a sequence of ``top hat'' field
configurations, parametrized by a single degree of freedom $\phi \left(
t\right) $, representing the field amplitude within a domain of size
$M^{-1}$,
outside of which the system field vanishes. The center of mass coordinates
of the ``hat'' may be treated as collective coordinates in the usual way,
and do not affect our results \cite{Rajaraman}.

If the quantum state of the full field is described by a density matrix $%
\rho (\phi ,\varphi ,\phi ^{\prime },\varphi ^{\prime },t)$, the state of
the $\phi $ field is described by the reduced density matrix $\rho
_{r}\left( \phi ,\phi ^{\prime },t\right) =\int d\varphi \;\rho \left( \phi
,\varphi ,\phi ^{\prime },\varphi ,t\right) \text{,} $ or equivalently by
the reduced Wigner function 
\begin{equation}
f\left( \phi ,p,t\right) =\int \frac{du}{2\pi \hbar }\;e^{-ipu/\hbar
}\,\rho
_{r}\left( \phi +\frac{u}{2},\phi -\frac{u}{2},t\right) \text{.}
\end{equation}

To second order in $g$ and leading order in $\hbar $, $f$ evolves according
to \cite{paper2} 
\begin{equation}
\frac{\partial f}{\partial t}=\{H_{s},f\}+\frac{\partial }{\partial
p}\left(
\Gamma f+ \hbar \left\{ N,f\right\} \right)
-\frac{g\hbar ^{2}}{24}\frac{\partial^{3}f}{\partial p^{3}}\text{,}
\label{lange4}
\end{equation}
where the curly brackets are Poisson brackets, $H_{s}=\dot{\phi}%
^{2}/2+V\left( \phi \right) $, $\Gamma =\int dt^{\prime }H\left(
t-t^{\prime
}\right) \phi \left( t^{\prime }\right) ,\;$and $N=\int dt^{\prime }N\left(
t-t^{\prime }\right) \phi \left( t^{\prime }\right) .$ The kernels $H$ and
$N
$ represent the effects of dissipation and noise, respectively. They come
from the quadratic part of the Feynman-Vernon influence action \cite{CH89}.
Computing the influence functional requires handling formally infinite
quantities (and in our case, also a linear term in $\phi $);
regularization and renormalization leave a finite residuum, which are the
one-loop correction to the effective potential and a finite wave function
renormalization. As we have already remarked, we assume that these
corrections are already included in Eq. (\ref{action}). This is sensible
because, although they may be quantitatively important, they do not affect
the nature of the problem \cite{prefactor}.

Our approach to Eq. (\ref{lange4}) will be the following. It is clear that
if only the first term of the right-hand side is kept, the equation reduces
to the classical transport equation and there is no tunneling. Retaining
the
first and last term on the right-hand side is equivalent to writing a
Schr\"{o}dinger equation for the wave function of the homogeneous mode, as
if it were a closed system. Since this is a one-dimensional problem, the
tunneling rate may be computed either by the instanton or the WKB method,
which are known to be equivalent in this case. We wish to know if the
middle
(back-reaction) term makes a substantial contribution to the total rate.
With this strategy in mind we shall discard the third contribution to the
right-hand side in Eq. (\ref{lange4}), assuming implicitly that the
back-reaction term is dominant (see below). It then reduces to Kramers'
equation which may be seen \cite{paper2} to describe the evolution of an
ensemble of points evolving according to the Langevin equation: $d\phi
/dt=p(t),$ 
\begin{equation}
\frac{dp}{dt}\left( t\right) =-V^{\prime }\left[ \phi \left( t\right)
\right] +\int dt^{\prime }\;H\left( t-t^{\prime }\right) \phi \left(
t^{\prime }\right) +\xi \text{,}  \label{lange2}
\end{equation}
with initial conditions $(\phi _{i},p_{i})$ weighted according to the
initial Wigner function, and Gaussian noise characterized by $\langle \xi
(t)\xi (t^{\prime })\rangle =\hbar N(t-t^{\prime })$. Although this
representation of the dynamics has an important heuristic value, only the
reduced Wigner function $f$ has a direct physical meaning, and it will not
allow an interpretation as a classical distribution function in general (it
will not be generally positive definite). 

We are interested in the weak dissipation limit, as discussed by Kramers 
\cite{Kramers}, when the relaxation time is long compared with the
classical
period of motion. On the other hand, the memory time in the integrals in
Eq.
(\ref{lange4}) is determined by the frequencies in the environment, which
are large with respect to the dominant frequencies in the system. Thus we
are allowed to (and, in a formal expansion in powers of $\hbar $, we must)
use solutions to the classical equations of motion within the memory terms;
these solutions may be written down explicitly in terms of elliptic
functions. This is a less drastic approximation than the Markovian one
discussed in Ref. \cite{Berera}. We also neglect transient terms (or in
other words, we assume $t\gg M^{-1}$); this means that we can take the
lower
limit of the time integrals in Eqs. (\ref{lange4}) as $t=-\infty $. In the
weak dissipation limit the reduced Wigner function depends only on the
action variable $J=\frac{1}{2\pi }\oint d\phi \;p$, and, averaging over
angles, Kramers' equation reduces to a one-dimensional Fokker-Planck
equation $\partial f / \partial t + \partial \bar{\Phi} / \partial J=0$,
where $\bar{\Phi}=-\Theta \Omega^{-1} \partial f / \partial J %
-\Lambda f$ is the flux, and $\Omega =\Omega \left( J\right) $ is the
frequency of the corresponding classical motion \cite{Risken}. The point of
this analysis is that it is possible to derive explicit expressions for the
coefficients $\Theta $ and $\Lambda $ in the Fokker-Planck equation \cite
{CV,paper2}. Near the value $J_{s}$ of the action variable at the
separatrix
(that is, the limiting trajectory which connects to the unstable
equilibrium
point) these have finite values, while as $J\rightarrow 0$ they go to zero
as $E^{3}$, where $E=H_{s}\left( J\right) $. In the thermal activation
problem one finds an identical equation, but the coefficients decay
linearly
on $E$ \cite{Kramers}.

The weakness of noise and dissipation in our (vacuum decay) problem
reflects
the origin of these effects in particle creation. Since particles are
created in pairs, there is a threshold for particle creation at frequency
$%
\omega \sim 2M$. At low energy, classical motion is mainly harmonic with
frequency $M$ for small oscillations around the metastable minimum, hence
particle creation is weak. It never actually vanishes, though, because at
any finite energy there is a small deviation from harmonic motion. The
amplitude of the component with frequency $n\Omega $ decays as $E^{n}$ as
$%
E\rightarrow 0$, which is enough to trigger particle creation \cite{CH97}.

The Fokker-Planck equation describes an initial value problem subject to
nontrivial boundary conditions at $J=0$ and $J=J_{s}$. These are vanishing
flux $\bar{\Phi}=0$ at $J=0$, and vanishing probability $f=0$ at $J=J_{s}.$
The linear operator $L$ which is defined by $Lf=\partial
\bar{\Phi}/\partial
J$ is self-adjoint with respect to an adequate inner product \cite{paper2},
and the equation may be solved by an expansion in normal modes in the usual
way. A general solution is reconstructed as a superposition of modes
$f_{r}$
decaying as $\exp \left( -rt\right) $. For a given $r$, $f_{r}$ oscillates
as $J\rightarrow 0$, and the modes must be subject to a continuum
normalization, as in the usual treatment of the WKB\ wave function in
quantum mechanics \cite{Landau}. The result is that, given any smooth
initial condition with mean energies of the order of the false vacuum
energy 
$\hbar M/2$, the persistency amplitude $P\left( t\right) =2\pi \int
dJ\;f\left( J,t\right) $ decays exponentially with a constant $\lambda $
for 
$\lambda t\gtrsim 1$, turning to $1/t$ for longer times (this crossover is
also observed in the usual tunneling amplitude \cite{Khalfin}). The
constant
is \cite{paper2} $\lambda \approx \Delta \exp \left\{ -\int dE\;\frac{%
\Lambda }{\Theta }\left( E\right) \right\} \sim \Delta \exp \left\{
-a\frac{%
M^{2}}{\hbar g^{2}}\right\} $, where $\Delta $ is of order $1$, and $a\sim
0.2$. By contrast, the tunneling amplitude, in the corresponding
approximation of only considering the homogeneous mode, yields a similar
formula, but with $a\sim 4.8$ \cite{Coleman}. We can see that, in this
case,
the zero temperature activation rate is higher than the tunneling amplitude
by an order of magnitude in the exponent.

We should point out that the fact that in our example the activation
amplitude is actually larger than the tunneling amplitude is model
dependent. Roughly speaking, low and broad barriers favor activation, while
high and narrow barriers favor tunneling. We must also stress that these
results must be considered as preliminary, pending a more satisfactory
parametrization of the {\it system}, and therefore a more realistic
modelization of the system-bath interaction. It is safe to conclude,
however, that activation should not be discarded {\it a priori, } but
rather
counted on as a potentially significant contribution to the overall decay
amplitude.

Note that we reach a different conclusion to that of Ref. \cite{Caldeira},
where dissipation suppresses tunneling. However a direct comparison between
the two analysis cannot be done as the two models differ essentially in the
coupling with the environment degrees of freedom, which is quadratic in our
case. Our analysis is closer to that of Ref. \cite{Bak}, and it is
certainly
compatible with their conclusions; see also Ref. \cite{Fujikawa}.

Still, we must stress that we should not expect a similar behavior in
systems with few degrees of freedom. The fact that in our problem the
environment actually contained a large enough number of degrees of freedom
as to represent a continuum for all practical purposes is essential to
provide a suitable driving force. If some frequency intervals were lacking,
then there would arise islands of stability where no resonance is strong
enough to move the system forward. These islands would act as absolute
barriers to noise-induced decay, or at least would depress the noise
induced
amplitude much below the tunneling estimates.

In conclusion, we have shown that vacuum decay in field theory is
qualitatively different from the same process in systems with few degrees
of
freedom, because the former are intrinsically open systems. Interaction
between long and short-wavelength modes induce a stochastic dynamics for
the
former and results in activation even at zero temperature.

\acknowledgements

We thank Leticia Cugliandolo, Bei-Lok Hu, Jaume Garriga, Rodolfo Id Betan
and
Jorge Kurchan for many illuminating discussions.
This work has been partially supported by Fundaci\'{o}n Antorchas under
grant A-13622/1-21. E. C. acknowledges support from Universidad de Buenos
Aires, CONICET, Fundaci\'{o}n Antorchas and the ANPCYT through project
PICT99 03-05229. A. R. and E. V. have also been supported by the CICYT
Research Project No. AEN98-0431. E.V. also acknowledges support from the
Spanish Ministery of Education under the FPU grant PR2000-0181 and the
University of Maryland for hospitality.


\end{document}